# Probabilistic genotyping code review and testing

John Buckleton D.Sc.[1,2], Jo-Anne Bright Ph.D.[1], Kevin Cheng M.Sc.[1], Duncan Taylor Ph.D., Ph.D.[3,4]

1. Institute of Environmental Science and Research Limited, Private Bag 92021, Auckland, 1142 New Zealand
2. University of Auckland, Department of Statistics, Auckland, New Zealand
3. School of Biological Sciences, Flinders University, GPO Box 2100 Adelaide SA, Australia 5001
4. Forensic Science SA, GPO Box 2790, Adelaide, SA 5000, Australia

**Abstract**

We discuss a range of miscodes found in probabilistic genotyping (PG) software and from other industries that have been reported in the literature and have been used to inform PG admissibility hearings.

Every instance of the discovery of a miscode in PG software with which we have been associated has occurred either because of testing, use, or repeat calculation of results either by us or other users. In all cases found during testing or use something has drawn attention to an anomalous result. Intelligent investigation has led to the examination of a small section of the code and detection of the miscode.

Previously, three instances from other industries quoted by the Electronic Frontier Foundation Amicus brief as part of a PG admissibility hearing (atmospheric ozone, NIMIS, and VW) and two previous examples raised in relation to PG admissibility (Kerberos and Therac-25) were presented as examples of miscodes and how an extensive code review could have resolved these situations. However, we discuss how these miscodes might not have been discovered through code review alone. These miscodes could only have been detected through use of the software or through testing. Once the symptoms of the miscode(s) have been detected, a code review serves as a beneficial approach to try and diagnose to the issue.

**KEYWORDS**

Probabilistic genotyping, forensic DNA analysis, mixtures, STRmix™, validation, code review.

**HIGHLIGHTS**

- A range of miscodes are discussed.
- The mode of discovery of each is reported.
- In all cases the miscodes were discovered by testing or use.
- There is no instance in this set of discovery by code review.
- There is no evidence that open source software is beneficial.

**Introduction**

There is a widespread interest in methods to independently test the operation of probabilistic genotyping (PG) algorithms that are used in forensic DNA testimony. This is completely reasonable given the importance that may be attached to the results of an analysis.

The usual method advocated for this test by the defense is independent review of the source code, or code review [1-4]. In the U.S., the defendant has a constitutional right to confront the witness against him [5, 6]. Wexler [7] and Matthews et al. [2] argue strongly for the extension of this right to code review. Cybergenetics, a developer of PG software within the US, maintains a list of court decisions regarding code review on its webpage [8]. The decisions on this list overwhelmingly deny the defendant access to the source code with the exception of New Jersey v Pickett [4] which is still being litigated.

In the development of any program, code review by the programmers is continuously happening. During this process, not only do the programmers review the code, but they will also test to ensure that it is functioning as expected. If a particular section or method of the code is tested and it is not functioning as expected or producing errors, then this is known as a software bug[1]. Programmers will attempt to debug this unexpected result. This can be done by first understanding what is causing this bug, usually through identifying the line or section of the code that is causing the issue. Once identified, a correction is attempted, then the programmers will test again. If software bugs are still present, then this cycle is continued until the code is correct. The testing of the software strives to cover all possible scenarios; however, users may operate the software in a different way than intended. This can result in software bugs in the postproduction of the software.

Here within the context of this paper, we refer to code review as a formal review by an independent individual, knowledgeable in computer coding, being engaged to review the source code of a PG software. There usually are time constraints imposed by the cost of payment to the reviewer and there may be logistical constraints such as supervision of the review. The belief is that a code review under these conditions is effective in finding software bugs. We change here to using the preferred term "miscodes" instead of the misleading term "bugs".

Our opinion is that code review alone is an ineffective solution to identify miscodes in PG software. It is through a combination of extensive testing of the PG software and code review that miscodes are identified and resolved. In this work we list the miscodes in probabilistic genotype software with which we have been associated and how they were discovered. We are comprehensive with regard to the software STRmix™ and definitely not comprehensive with regard to Lab Retriever, Forensic Statistical Tool (FST), and EuroForMix (EFM). We also discuss five software miscodes in other industries from research of others' work each of which makes a special point. This is a non-random sample but does give some insight into how miscodes are best found.

We also discuss the code reviews under non-disclosure agreements and end with some suggestions for forward movement.

1. **Some Miscodes in Probabilistic Genotyping Software**

In this section, we list a few of the miscodes identified in postproduction probabilistic genotype software and how they were discovered. This, by no means, is an exhaustive list of miscodes, as miscodes can occur during the routine development software which are quickly identified and corrected.

The pattern of discovery is the same in the 18 postproduction miscodes within various PG software described here. Attention is drawn to an example that is giving an unexpected result. This could be in use by a laboratory or by testing or experimentation. After

---

[1] We are asked to avoid the use of the word "bug." The term "bug" comes from none other than Thomas Edison who used it in an 1878 letter to an associate. He noted: "*You were partly correct, I did find a 'bug' in my apparatus, but it was not in the telephone proper. It was of the genus 'callbellum.' The insect appears to find conditions for its existence in all call apparatus of telephones.*" *Callbellum* is not a real insect.

The reason for avoiding the term is that it implies some external agent has caused the fault rather than making a candid statement that it was caused by human error.

investigation, often of intermediate calculations, attention is focussed on a small section of code. The miscode is then detected.

*1.1 STRmix™*

STRmix™ maintains a comprehensive list of postproduction faults that have been discovered on its webpage. Of the 14 postproduction miscodes found in STRmix™[2] three were noticed in use (two by users and one by the STRmix™ team) where unusual results were detected and investigated by the STRmix™ team. Eleven were detected by parallel calculation of intermediate results (three by one user and eight by the STRmix™ team). The three found by a user were identified during their internal validation process. The eight found by the STRmix™ team were often associated with developmental validation of a successor version but existed in earlier versions. None of the miscodes have been found by code review. None have been found as part of a judicial process.

The second miscode that was discovered in STRmix™ has featured prominently in discussion in admissibility for STRmix™ and in a code access case for TrueAllele® [9].

We were initially contacted by Queensland Health (Australia) in December 2014 who noticed that the *LR* they obtained from a single run of STRmix™ was slightly different from the *LR* obtained by a database search to the same individual. The specific case that triggered the query involved the deconvolution of a three-person mixture with one assumed contributor. The version of STRmix™ in use at Queensland Health at the time was V2.05. The fault would occur rarely under a specific set of circumstances. Its occurrence is exacerbated by over-estimating the number of contributors to a profile. This was a behaviour that we had inadvertently encouraged by giving the false suggestion that it was safer to overestimate the number of contributors than underestimate.

The miscode was corrected in STRmix™ V2.06 which was programmed, tested, and released within two weeks of the issue being raised by Queensland Health.

This miscode rarely caused a problem in most of Australia and New Zealand. In Queensland 23 cases resulted in a minor change to the numerical value of the likelihood ratio and new statements were issued. Thirty-seven (37) cases did not result in a change to the likelihood ratio that was reported in the original statement. In part this is because Australia caps reported *LR*s at 100 billion ($10^{11}$). We do not have available the numerical changes for Queensland but the magnitude of the change for different mixtures processed by us is given in Figure 1.

---

[2] https://strmix.com/news/summary-of-miscodes/

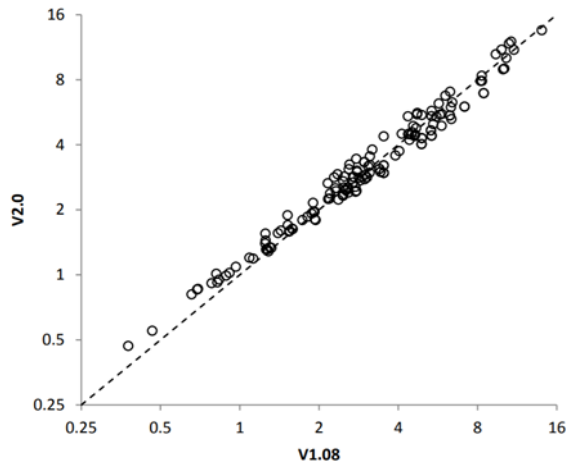

Figure 1: A plot of the locus log*LR*s from five two-person and five three-person Identifiler™ mixtures with a range of input DNA. This gives 150 comparisons in V2.06 (miscode corrected) versus V1.08 (miscode present).

The specific situation had not been envisaged and was not tested (or possibly tested but no effect noticed) at developmental or internal validation elsewhere.

Because the magnitude of the effect is small and intermittent it is close to impossible to detect this fault by observation of the output. We highlight this example because of the aspect that we do not think it could be detected by code review and is likely to be missed by testing. It was detected by an observant user. Other than this method of noticing it could only be detected realistically by repeat calculation of the locus *LR*s in an affected situation.

### *1.2 Lab Retriever*

A list of coding issues within the semi-continuous PG software Lab Retriever is maintained at GitHub[3].

Two miscodes were found by us in conjunction with Keith Inman, one of the Lab Retriever developers. These were found when we tried to replicate the *LR* for the simplest possible situation, a single-source profile with no dropout or drop-in. The miscodes were some entries in the allele probability files provided with the software and a parsing fault that meant that any value for probability of dropout less than 5% was treated as 5%. We note that these miscodes were immediately rectified and tended to have a small numerical effect on the *LR*.

Noting that Lab Retriever is open source and that these two faults, the only two that involved us, were found by the simplest test possible we observe in this instance:

1. The value of testing even of the simplest type, and
2. The fact that no-one had either looked at the source code or noted the errors.

### *1.3 EuroForMix (EFM)*

EuroForMix (EFM) is an open-source PG software. The primary developer is Øyvind Bleka. The source code and a list of version changes for each released version of EFM can be found on GitHub[4]. Version change information for earlier versions of EFM can be found on the EFM website[5]. In the list of changes, EFM discloses which function and line number(s) had a miscode. A list of miscodes could be compiled from this. We have been responsible for finding two miscodes in EFM, both through testing the software. Here, we described one in detail.

---

[3] https://github.com/SCIEG/LabRetriever/issues
[4] https://github.com/oyvble/euroformix/blob/master/NEWS

As part of experimental work unconnected with any validation a team of some of the STRmix™ and the EFM developers began a comparison of the *LR*s obtained by EFM v3.0.3 and STRmix™ v2.7.0 on a set of one- to four-person mixtures. A comparison of *LR*s from the two software for the four-person mixtures is given in Figure 2.

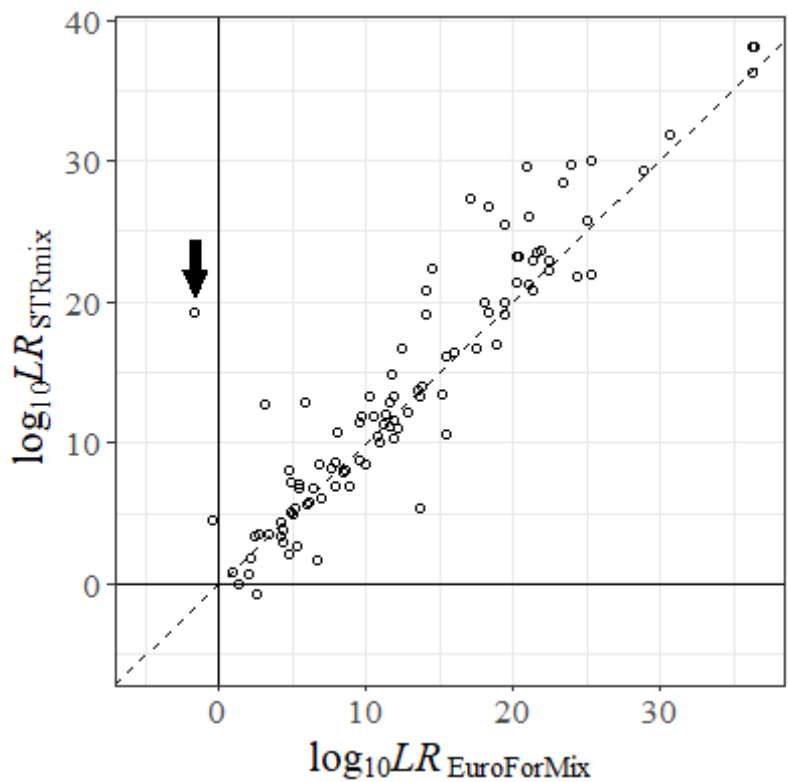

Figure 2: Scatter plot of the STRmix™ $\log_{10}$LR and EuroForMix $\log_{10}$LR for known contributors to four-person mixtures. The datum examined in detail (PROVEDIt H09_RD14-0003-48_49_50_29-1;4;4;4-M2a-0.75GF-Q0.4_08.25sec) is marked with an arrow.

One analysis from the four-person mixtures (PROVEDIt H09_RD14-0003-48_49_50_29-1;4;4;4-M2a-0.75GF-Q0.4_08.25sec) that showed a substantial difference in the $\log_{10}LR$ between EuroForMix (-1.75) and STRmix™ (19.79). A locus-by-locus comparison found that the largest discrepancy by far was at SE33 (EFM -5.77 versus STRmix™ 2.19) but we also look at the FGA locus (EFM 0.17 versus STRmix™ 2.67). These loci are shown in Figure 3.

---

[5] http://euroformix.com/changes

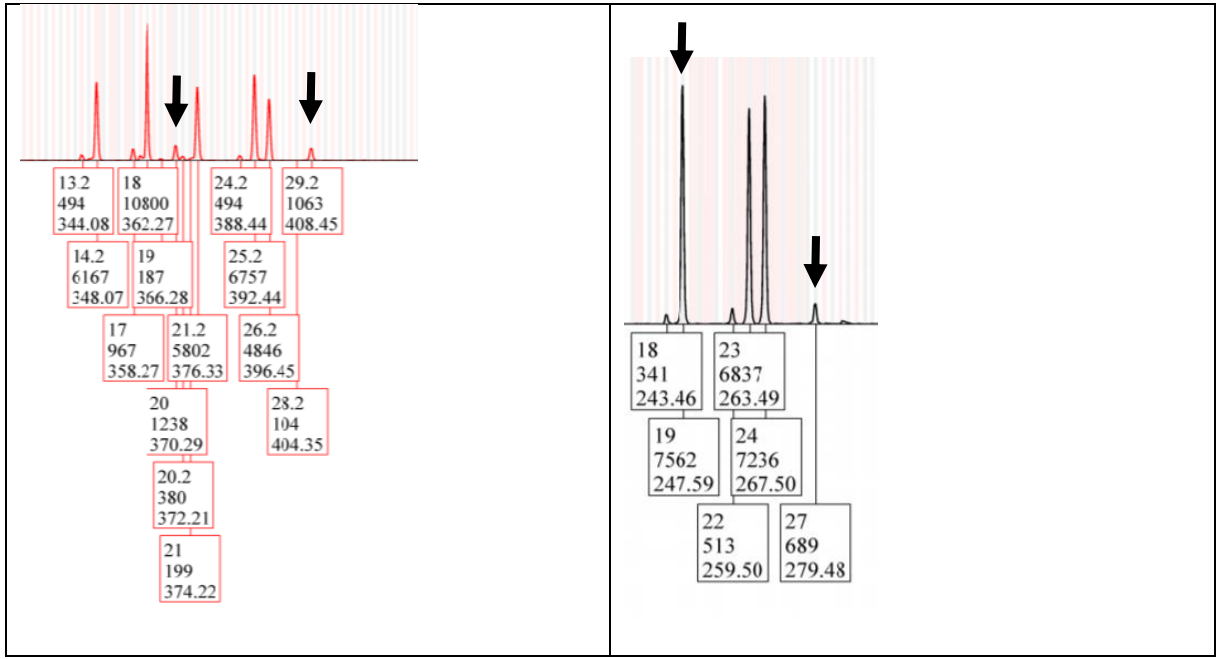

Figure 3: The SE33 locus (left) and FGA locus (right) from four-person mixture PROVEDIt H09_RD14-0003-48_49_50_29-1;4;4;4-M2a-0.75GF-Q0.4_08.25sec. This is targeted as a 4:4:1 mixture. The black arrows indicate the ground truth minor.

SE33 did not seem too problematic to deconvolute. Human inspection can readily identify the minor as 20,29.2 (the peak heights are 1238 and 1063 rfu respectively). STRmix™ places 100% of its weight on the 20,29.2 minor and this is correct.

There has been much talk that there is no correct *LR*. In one, fairly impoverished, sense this is true. However, we can state unambiguously that a log*LR* of -5.77 for the minor contributor for this locus is incorrect. In fact, anything below 0 is obviously incorrect. Further, a fully resolved minor 20,29.2 gives $\log_{10} LR = 2.19$. Therefore, we can state that 2.19 is the correct result given acceptance of the models underlying the calculation. We could still discuss whether the models are correct—they will only ever be approximations—but at least for this situation, 2.19 is the result desired of the software.

The FGA locus also seems straightforward. The minor is 19,27. One allele of the minor can be clearly seen at the 27 position. The other minor allele cannot be placed with confidence but should most probably be masked under the 19, 23, or 24. We could even substantiate that there should be a small preference for the minor allele being masked under the 19 allele. However, without making this assessment and simply stating that the other minor allele should be masked somewhere suggests $\log_{10} LR$ for this locus of 2.55. We feel that it can be concluded that the EFM $\log_{10} LR$ for this locus of 0.17 can be considered incorrect.

In this case, therefore, our attention was drawn to one mixture, then one locus, and we could assign right, and hence wrong answers. Given this information by EFM's developer, Øyvind Bleka, who was intimately familiar with the code, then examined the code and diagnosed the miscode (which was immediately announced and corrected[6]). EFM had not found the minor

---

[6] **Message 30. March 2021:** Relevant for EuroForMix version v3.0.0-v3.2.0

- A bug was recently discovered when all the alleles in the defined allele-frequency table are also observed in the evidence profile (for one of the markers), and at the donor genotype even though we suggest that it is straightforward for a software or a human to do so. We think we could have discovered it in testing alone without the comparison with STRmix™ since the false exclusion would have attracted our attention. The diagnosis was that this difference was due to a miscode. We needed help to understand this error as we could not find it ourselves in the code despite being pointed to the correct area. Bleka has assisted us with a pseudocode rendition of the incorrect and the corrected versions.

---

# Incorrect version

**Begin**

**for** the allele proposed at each locus store an index indicating which allele stutters on to it.

    always **append** a 'dummy-index' for each locus to take into account that last allele is the 'Q allele' [this is an allele that represents all alleles not seen in this mixture and is used when drop-out is proposed]

**End**

---

# Corrected version

**Begin**

**for** the allele proposed at each locus store an index indicating which allele stutters on to it.

    **if** a 'Q allele' is defined at a particular locus, **append** a 'dummy-index' for the corresponding locus to take into account that last allele is the 'Q allele' [this is an allele that represents all alleles not seen in this mixture and is used when drop-out is proposed]

**End**

---

EFM is open source. Even with help and being directed to the correct section of code we struggled to see the miscode and how it interacts with other aspects of the software. The miscode could have only been diagnosed by the developer. We offer the following:

1. This miscode was found by testing, and
2. It is very hard for an external party to see what the miscode is, even when directed to the exact spot.

The rapid production of a fix allowed a comparison of the affected and corrected *LR*s. In Figure 4 we give the *LR*s for 13 four-person mixtures from the PROVEDIt dataset [10] analysed in both versions. This led to *LR*s for 52 true donors and 3198 false donors. Nine of 52 true donor comparisons changed by more than three orders of magnitude. 718 of 3198 false donor comparisons also changed by more than three orders of magnitude.

---

    same time a stutter model (BW or/and FW) is assumed (full details provided in the attached document DiscoveredBug_EFM3_noQalleleStutter.pdf).
- Please note this bug is fixed in the released version EuroForMix v3.3.0, which also has some other updates included: It is recommended that users switch to this version.

http://www.euroformix.com/sites/default/files/DiscoveredBug_EFM3_noQalleleStutter.pdf

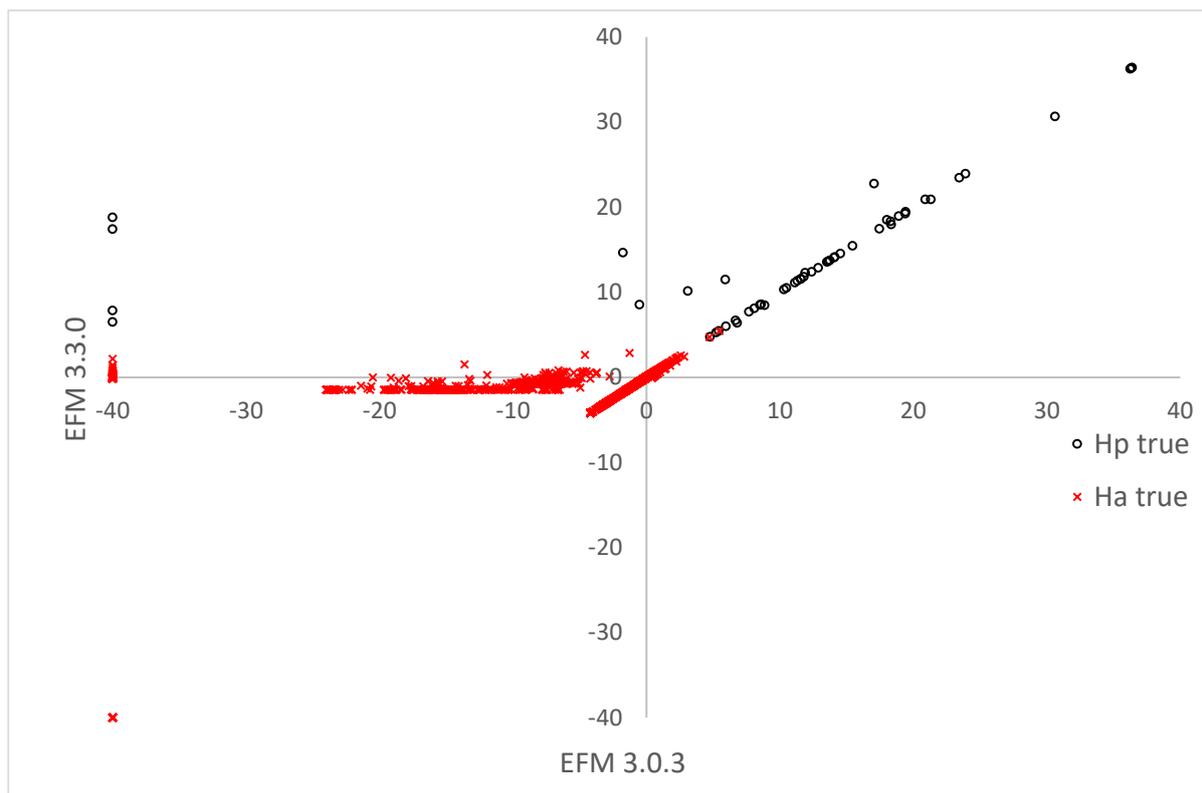

Figure 4: The *LR*s produced by EuroForMix V3.0.3 and V3.3.0 on the four-person PROVEDIt mixture set examined.

Although this miscode was detected in experimental work unconnected with validation, we do think that it could have been found in validation. It is worthwhile discussing the steps that would have been required. The problem appears to be limited to, or mainly limited to four-person mixtures. To be detected, obviously, a potentially affected mixture needs to be examined. First, the low $\log_{10}LR$ for this mixture needed to be noticed. We think that this is likely. Next the locus-by-locus $\log_{10}LRs$ needed to be examined and the low value for SE33 noticed. Again, we think this is likely. Last, inspection of the reference sample for the minor and the profile would indicate that the result was wrong.

Given the magnitude, frequency, and diagnosability of the error we ask the question: "Why was it not found during developmental validation?" We assume that the relevant four-person tests were either not done or if done were not examined sufficiently.

### *1.4 FST software*

Forensic Statistical Tool (FST) was developed for use by the New York City Office of Chief Medical Examiner (OCME), specifically to interpret DNA profiles processed in their laboratory [12]. It is no longer used for casework. FST had a function which removed a locus from the likelihood ratio calculation if the sum of the allele probabilities at that locus was greater than or equal to 0.97 [13] @ pg 172 ln 10. This rule was applied if this sum was reached in any of the four subpopulations used by FST (Asian, Black, Caucasian, and Hispanic).

This has been criticized, in part, because it was not mentioned in the publication or validation materials and the software does not alert the user to this rule being invoked. This function was discovered by defense analysts when rerunning some of the validation samples[7]. For one sample described as "pen B" they obtained a different *LR* from the one given in the validation document[14] @ pg 741. This was traced to the function that drops loci from the calculation[14] @ pg 742.

Omitting a locus effectively assigns an *LR* of 1 for that locus. Adams reports that the *LR* for these three loci should be D3 0.53, D13 3.13, and D16 1.33 [15] @ pg 4. The omission is therefore conservative (using the lower *LR* definition) for D13 and D16 and non-conservative for D3 in this case.

The first lines of the function are given below. The text after /// until the end of the line is comment and is not part of the functional code.

```
/// This function checks for the total frequencies according
to races and removes the allelles (sic) from calculation
/// if the sum of frequencies are greater than 0.97.
/// </summary>
public void CheckFrequencyForRemoval(DataTable dtFrequencies)
```

There was clearly no effort to conceal this function in the code however the code itself was not released for inspection until a court order.

Matthews et al. [2] report an empirical comparison. It is difficult to obtain the data we want from the publication since we need to read the details of graphs (their Figures 1 and 2) and summary tables (their Tables 1 and 2). It appears to us that the locus dropping function has:

1. Tended to lower the *LR* for true donors with *LR* > 1, and
2. Tended to raise the *LR*s for false donors.

This could be interpreted as conservative for true donors and non-conservative for false donors. Clearly this is an undesirable behaviour however the effect looks to be small.

Gasston et al. [16] find "*On average, the dropping of a locus is conservative for six-peak loci and nonconservative for five-peak loci. For persons of interest (POIs) with rare alleles, the dropping is usually conservative. For POIs with common alleles, the dropping of the locus is often nonconservative.*"

Based on this finding, we believe that the locus dropping function does not systematically assist the prosecution. The function is clearly signalled in the code and could have been found by code review. The fact that it was found by testing does not mean that it could not equally have been found by code review.

### 2. Other Miscode Examples

In an amicus brief Electronic Frontier Foundation (EFF) [3] state: "*The necessity of independent source code review for probabilistic DNA programs was starkly demonstrated when FST (a counterpart to STRmix that was used in New York crime labs) was finally provided to a defense team for analysis.*"

---

[7] We rely heavily on an email from Nathaniel Adams quoted here: "*When we ran the Pen B comparison on our copy of FST, we were concerned that it didn't match the LR reported in OCME's validation study or in the independent reproduction.*"

They quote the OCME FST undisclosed function and three further situations in support of their position. The OCME FST software was discussed above. The three other situations presented by the amicus brief relate to

1. the recording of ozone measurements in the stratosphere,
2. the National Integrated Medical Imaging System; and,
3. the Volkswagen omissions.

All of these are instances of miscodes discovered through testing. In this section, we will briefly discuss these situations and how they were discovered.

We will also discuss two other examples where miscodes – the *Kerberos 4 authentication protocol* and the *Therac-25 lethal accidents*. These examples have been discussed in part as supporting evidence for the code review of PG software; however, these situations were also discovered through testing.

### 2.1 Stratospheric Ozone

In 1978 Nimbus-7, a meteorological satellite, carried two new NASA sensors designed to measure the total amount of ozone in a given column of atmosphere [17]. These were the Solar Backscatter Ultraviolet (SBUV) instrument and the Total Ozone Mapping Spectrometer (TOMS). In October 1985 a British team of scientists using ground-based equipment found a significant reduction in ozone over Halley Bay, Antarctica. An inspection of the TOMS data found that it had detected, but not reported, a dramatic loss of ozone over Antarctica. Review of the TOMS data analysis software showed that it had been programmed to omit data that deviated greatly from expected measurements. As we understand, the initial TOMS readings, which also detected the significant reduction, were not reported and so did not set off any alarms. We note here that there does not appear to be any miscode at all, the software was functioning as intended and so no level of code review, or indeed software validation, would have uncovered an error. The root cause of the error here seems to be that an underlying assumption about the model of ozone depletion (specifically that it would occur in the upper stratosphere first and not make a dramatic difference to the entire column of atmosphere) was incorrect.

### 2.2 National Integrated Medical Imaging System (NIMIS)

The National Integrated Medical Imaging System was implemented in 2008. It captured and stored Radiology, Cardiology and other diagnostic images electronically. The issue related to the "<" symbol not being transferred to downstream applications. The NIMIS "<" symbol final report gives an example: "*There is < 50% stenosis noted in the Internal Carotid Artery*" would display incorrectly as "*There is 50% stenosis noted in the Internal Carotid Artery*".

This had the potential to cause patient harm [18]. However after review there actually were no instances of patient harm as a result of this error.

This issue was identified by a Consultant Radiologist in a Hospital who notified the NIMIS programme on 24th July 2017.

### 2.3 Volkswagen omissions

The US Environmental Protection Agency (EPA) [19] found that Volkswagen (VW) had intentionally programmed turbocharged direct injection (TDI) diesel engines to activate their emissions controls only during laboratory emissions testing. This caused the vehicles' $NO_2$ output to meet US standards during regulatory testing, while they emitted up to 40 times more $NO_2$ in real-world driving.

West Virginia University research Assistant Professor Arvind Thiruvengadam was hired to do a standard emissions tests on diesel cars in the U.S. Thiruvengadam tested two VW cars and found that the claims of low emissions were unjustified [20].

A team, led by Kirill Levchenko, a computer scientist at the University of California San Diego [21] obtained copies of the code running on Volkswagen onboard computers from the company's own maintenance website and from forums run by car enthusiasts. … "*We found evidence of the fraud right there in public view*," Levchenko said.

The specific piece of code was labelled the "acoustic condition" which detected the conditions occurring during an emissions test and activated emissions-curbing systems, which reduced the amount of nitrogen oxide emitted.

### 2.4 Kerberos 4 authentication protocol

The Kerberos authentication protocol is a widely published and used piece of security software. Kerberos is open source. A pseudorandom number generator (RNG) is seeded each time a session key is generated. Pseudorandom number generators are insufficiently random for secure keys because they are too deterministic and predictable. This predictability makes guessing keys from a pseudorandom number generator much easier than attempting to naively brute force the entire key space. In the case of Kerberos Version 4, keys can be electronically guessed in seconds, allowing an attacker to make use of the key and subvert the Kerberos authentication system.

The problems in the routine producing the random keys were identified in 1988. In 1989 a new RNG code was inserted into the Kerberos source tree but was never implemented. The issue was rediscovered in 1995 after the release of a software upgrade (version 5). Because the code had been checked and the RNG supposedly fixed in 1989, everyone assumed that it worked. No one checked the open-source code to see if the fix had been implemented.

This and the FST undisclosed function are the only clear cases within this paper where code review could have worked in the detection of a miscode.

### 2.5 Therac-25 lethal accidents

Between 1985 and 1987 a computer-controlled radiation machine, the Therac-25, massively overdosed six people, some fatally. An overview is given by Levison [22]. Adams et al. conclude that the Therac-25 failure "highlights the importance of careful, independent V&V (verification and validation) of software that performs critical functions" [23] such as PG.

The Therac-25 had a turntable that rotated accessory equipment into place to produce one of two modes: X-ray and electron mode. The raw electron beam is highly damaging to human tissue and needs the beam energy and current moderated and the beam spread by magnets. After extensive testing a physicist and one of the operators were able to reproduce the conditions that led to the fault.

There are a very large number of causative factors of the fatal overdoses including general poor software design and development practices, and institutional complacency. We concentrate here on the race condition miscode.

A race condition is where the software's performance is dependent on the sequence or timing of other uncontrollable events. At least four miscodes were found in the Therac-25 software that could cause radiation overdose [24], the first two being race conditions. These were:

1. If the data were entered quickly the turntable could be left in the wrong position.
2. If the operator changes the beam type and power in the first 8s and moved the cursor to the final position, the system would not detect the changes.

The following is a very highly simplified pseudocode description of the key routine for the race condition in the Therac-25 [22] written so that it can be read by lay persons. We challenge the reader to find the miscode and note that we have concentrated the reader on the offending short section.

---

**Datent**: [this is the name of this subroutine}

**begin**
   calculate the table index
   call subroutine Magnet
   **if** mode/energy changed **then** return [exit this subroutine}
   [mode/energy changed is a shared variable set by the keyboard operator that indicates an editing request}
**end**

---

**Magnet**: [this is the Magnet subroutine}

   set the bending magnet flag
   repeat
     set the next magnet
     call Ptime [Ptime is a subroutine that introduces a time delay}
     **if** mode/energy changed **then** exit
   **until** all the magnets are set
   return [exit this subroutine}

---

**Ptime**: [this is the subroutine that adds the time delay}

   repeat
     **if** the bending magnet flag is set **then**
       **if** mode/energy changed **then** exit
   **until** the delay has expired
   clear the bending magnet flag
   return

---

There are two magnets in the Therac-25. The miscode is that the bending magnet flag is set at the start of Magnet and cleared on first exit of Ptime. On the second call of Ptime the bending magnet flag is now clear - we do not enter the **if** mode/energy changed **then** exit command.

We think the bug could be fixed by moving the command: "clear the bending magnet flag" command from Ptime to Magnet.

### 3. Discussions

Every instance of the discovery of a miscode in PG software with which we have been associated has occurred either because of testing, use, or repeat calculation of results either by us or other users. In all cases found during testing or use something has drawn attention to an anomalous result and intelligent investigation has eventually led to examining a small section of the code.

The three instances given by EFF (atmospheric ozone, NIMIS, and VW) and the Kerberos and Therac-25 incidents were all detected by testing or use. VW code was available by accessing the software onboard the car's computer and Kerberos had openly available code.

Open-source software have some perceived benefits, for example the source code is available to the community so that others can support and improve a solution. This is the case for open-source software with a large community or many contributors to the project. For example, TensorFlow, an open-source machine learning platform. Based on the GitHub page, it has over 150,000 users and 3000 contributors. Or the Linux kernel, a Unix-like computer operating system used by many institutes around the world is also an open-source project. However, in our experience, there is little evidence open source can lead to more reliable software. We suggest that this arises from the fact that code review is difficult by anyone other than the developer and that there is very little motivation for third parties to undertake a code review. This is possibly because the field of forensic mixture interpretation is more niche and has a smaller user-base. There is not a large community of subject-matter experts that are actively scrutinizing and evaluating the open-source code. We have had quoted to us sentences of the type "it's open source so anyone can check for themselves." The Kerberos example described above, and our own experience, suggests that open source may lead to the false expectation that the software has been extensively tested and reviewed.

We note that there has been a focus on code review as necessary to the defense of the accused [2, 7, 25]. We see discussion of whether code review as necessary or a right as misplaced; the issue is the conditions under which the code review happens. Some defence analysts require, possibly under court order, essentially unfettered access to the code at their own desk, with no time constraint, and no personal sanction for inadvertent code disclosure. Our own intellectual property lawyers would state that this is completely ineffective protection.

Our personal belief is that the defendant should have access to code review. We point out that this is unlikely to find a miscode without being coupled with testing but that it may highlight poor coding practice. We do not see value in debating whether this right should be respected, since we believe it unambiguously should, but we do see value in some, possibly mandated, consensus on the practicalities of how this can be achieved.

Similarly, the DNA Commission of the ISFG [26] "*does not consider examination of the source code to be a useful fact-finding measure in a legal setting. … However, if requested by the legal system, the code should be made available subject to the software provider's legitimate copyright or commercial interests being safeguarded. Supervised access to the code under a "no copy" policy is acceptable.*" As the developers of STRmix™ we also support the right of the accused to access the source code. To preserve the intellectual property this is done under a non-disclosure agreement and supervision [27].

In our own experience and discussed within this paper, code review without being in conjunction with extensive testing, is unlikely to discover a fault. It can, however, highlight poor coding practice. We further note that it is safe to place an executable file in the possession of the defense and this would allow as much testing as is desired or could be afforded. STRmix™ makes an executable file available to the defense for 60 days renewable. During this time limited access expert witnesses not only have access to the software to test many different scenarios, but they will also have access to the manuals to get a better understanding on how the software is expected to perform.

**Acknowledgements**

This work was supported in part by grant NIJ 2020-DQ-BX-0022 from the US National Institute of Justice. Points of view in this document are those of the authors and do not necessarily represent the official position or policies of their organizations.